\documentclass[fleqn,12pt]{olplainarticle}

\usepackage{microtype}
\usepackage{xspace}
\usepackage{graphicx}
\usepackage{xcolor}
\usepackage{float}
\usepackage{subcaption}
\usepackage{booktabs} 
\usepackage{hyperref}
\usepackage{amssymb}
\usepackage{amsmath}
\usepackage{gensymb}

\usepackage[T1]{fontenc}
\usepackage[misc,geometry]{ifsym} 

\makeatletter 
\renewcommand\@biblabel[1]{#1.}
\makeatother 

\title{Reducing false-positive biopsies with deep neural networks that utilize local and global information in screening mammograms}

\author[1]{Nan Wu}
\author[1]{Zhe Huang}
\author[1]{Yiqiu Shen}
\author[2]{Jungkyu Park}
\author[1]{Jason Phang}
\author[1]{Taro Makino}
\author[2,3,4]{S. Gene Kim}
\author[1,5,6]{Kyunghyun Cho}
\author[2]{Laura Heacock}
\author[2,3]{Linda Moy}
\author[2,4,1]{Krzysztof J. Geras}
\affil[1]{Center for Data Science, New York University}
\affil[2]{Department of Radiology, New York University School of Medicine}
\affil[3]{Perlmutter Cancer Center, NYU Langone Health}
\affil[4]{Center for Advanced Imaging Innovation and Research, NYU Langone Health}
\affil[5]{Department of Computer Science, Courant Institute of Mathematical Sciences, New York University} 
\affil[6]{CIFAR Associate Fellow}


\begin{abstract}

Breast cancer is the most common cancer in women, and hundreds of thousands of unnecessary biopsies are done around the world at a tremendous cost. It is crucial to reduce the rate of biopsies that turn out to be benign tissue. In this study, we build deep neural networks (DNNs) to classify biopsied lesions as being either malignant or benign, with the goal of using these networks as second readers serving radiologists to further reduce the number of false positive findings. We enhance the performance of DNNs that are trained to learn from small image patches by integrating global context provided in the form of saliency maps learned from the entire image into their reasoning, similar to how radiologists consider global context when evaluating areas of interest. Our experiments are conducted on a dataset of 229,426 screening mammography exams from 141,473 patients. We achieve an AUC of 0.8 on a test set consisting of 464 benign and 136 malignant lesions. 

\end{abstract}

\begin{document}

\flushbottom
\maketitle
\thispagestyle{empty}

\section*{Introduction}

Breast cancer is the most common cancer in women worldwide, after skin cancers and about 42,170 women will die from breast cancer in the United States for 2020, according to The American Cancer Society's estimate~\cite{breastcancer}. Screening mammography, a low-dose X-ray examination, is typically used for early detection of breast cancer. The United States Preventive Services Task Force suggests women urdergo such exams every two years if they are 50 to 74 years old and are at average risk for breast cancer~\cite{jama2019}. Although multiple studies have demonstrated that screening mammography reduces breast cancer mortality~\cite{lees2010theoretical, 10.1001/jama.2015.12783, jama2019, Monticciolo}, performance benchmarks demonstrate that 10\% of the performed exams are recalled for additional imaging, and approximately 80\% of biopsies subsequently performed are benign~\cite{lehman2017national}. The yearly national cost of breast-care caused by the false positive mammograms is estimated to be several billion of dollars~\cite{vlahiotis2018analysis, ong2015national} and for women with a false positive diagnosis, their mean cost of breast-care is even higher than the cost of breast-care services for women with cancer~\cite{chubak2010cost}.  It is therefore an important task to reduce the recall and biopsy rates so that to decrease patients’ anxiety and reduce healthcare costs while still maintaining optimal cancer detection rates, according to relevant guidelines~\cite{10.1001/jama.2015.12783}.
 
\begin{figure}[h!]
    \centering
    \begin{tabular}{ll}
        \includegraphics[width=0.25\textwidth, clip, trim=0mm 0mm 0mm 0mm]{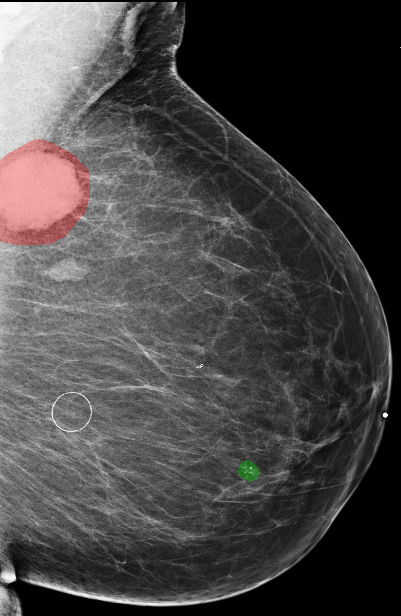} &  \includegraphics[width=0.25\textwidth, clip, trim=0mm 10mm 0mm 10mm]{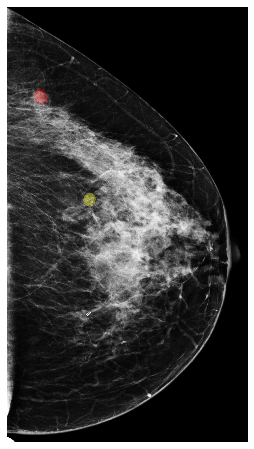} \\
        \textbf{a} & \textbf{b}
    \end{tabular}
    \caption{Images with both malignant lesion and benign tissue. \textbf{a}. An image of the left breast from mediolateral oblique view (L-MLO). The breast has two lesions confirmed by biopsy, one as malignant (annotated with red), and the other as benign (annotated with green). \textbf{b}. An L-MLO mammogram image from another patient. There are two lesions on the image, one as malignant (annotated with red), and the other as high-risk benign (annotated with yellow).}
    \label{fig:multi_lesion}
\end{figure}

Traditional computer-aided detection (CAD) tools for mammography neither detected more breast cancers nor decreased the recall rates for additional imaging~\cite{lehman2015diagnostic, fenton2007influence}. Early studies used deep neural networks (DNNs) to assist radiologists interpreting screening mammograms by making predictions for cancer of each breast~\cite{zhu2017deep, kyono2018mammo, aboutalib2018deep, kim2018applying, wu2019deep, shen2020interpretable, mckinney2020international}. This task is frequently considered in literature. It can be viewed as breast-level classification, and models developed accordingly have shown comparable performance to radiologists~\cite{wu2019deep, shen2020interpretable, mckinney2020international}. However, these models suffer from performance degradation when evaluated on a population only containing exams which lead to biopsies, without healthy breasts as negative cases~\cite{wu2019deep}. Meanwhile, models built for the breast-level classification task cannot provide independent risk estimations for multiple areas of interests appearing in the same breast. It is common to encounter cases with multiple findings~\cite{cohen2020multiple}. For example, multiple bilateral circumscribed breast masses are detected in approximately 1.7\% of routine screening mammograms~\cite{leung2000multiple}. In the NYU Breast Cancer Screening Dataset~\cite{wu2019nyu}, a representative sample of screening mammograms from 2010 to 2017, there are 7.45\% images with more than one annotated lesions, and 25.75\% of these images have lesions of different categories. Some examples are shown in Figure~\ref{fig:multi_lesion}. In light of this, the previously proposed models for breast-level classification are difficult to use for the goal of reducing unnecessary biopsies.

Besides breast-level malignancy classification, deep learning methods have also been used to identify high-risk lesions~\cite{ribli2018detecting, liu2020cross, samala2016mass, agarwal2019automatic}. Some of these works can provide risk estimation across regions of the breast, but usually only consider information in a small local region~\cite{samala2016mass, agarwal2019automatic}. The majority of the existing works often utilize object detection models such as Mask-RCNN~\cite{he2017mask}, which neither explicitly utilize fine details nor consider global context. In contrast, radiologists often consider global context factors to make their diagnoses~\cite{wei2011association,pereira2009spatial}. These global context factors include the mammographic breast density, i.e. the global amount of fibroglandular tissue, and the associated parenchymal and nodular patterns of the breasts~\cite{pereira2009spatial}. Dense fibroglandular breast tissue is a known risk factor for breast cancer~\cite{10.1001/jamaoncol.2018.7078}. Other global context factors include the distribution of microcalcifications in the tissue adjacent to an index lesion, or throughout the breast. These global findings often affect radiologists' level of suspicion for a particular lesion. Within deep learning methodology, these scenarios could be viewed as utilizing global image context for classifying a patch of an image. This motivates investigating whether global context is as important for neural networks as it is for human experts.

\begin{figure}[h!]
    \centering
    \begin{tabular}{ccc}
        \includegraphics[width=0.7\textwidth, clip, trim=0mm 0mm 0mm 0mm]{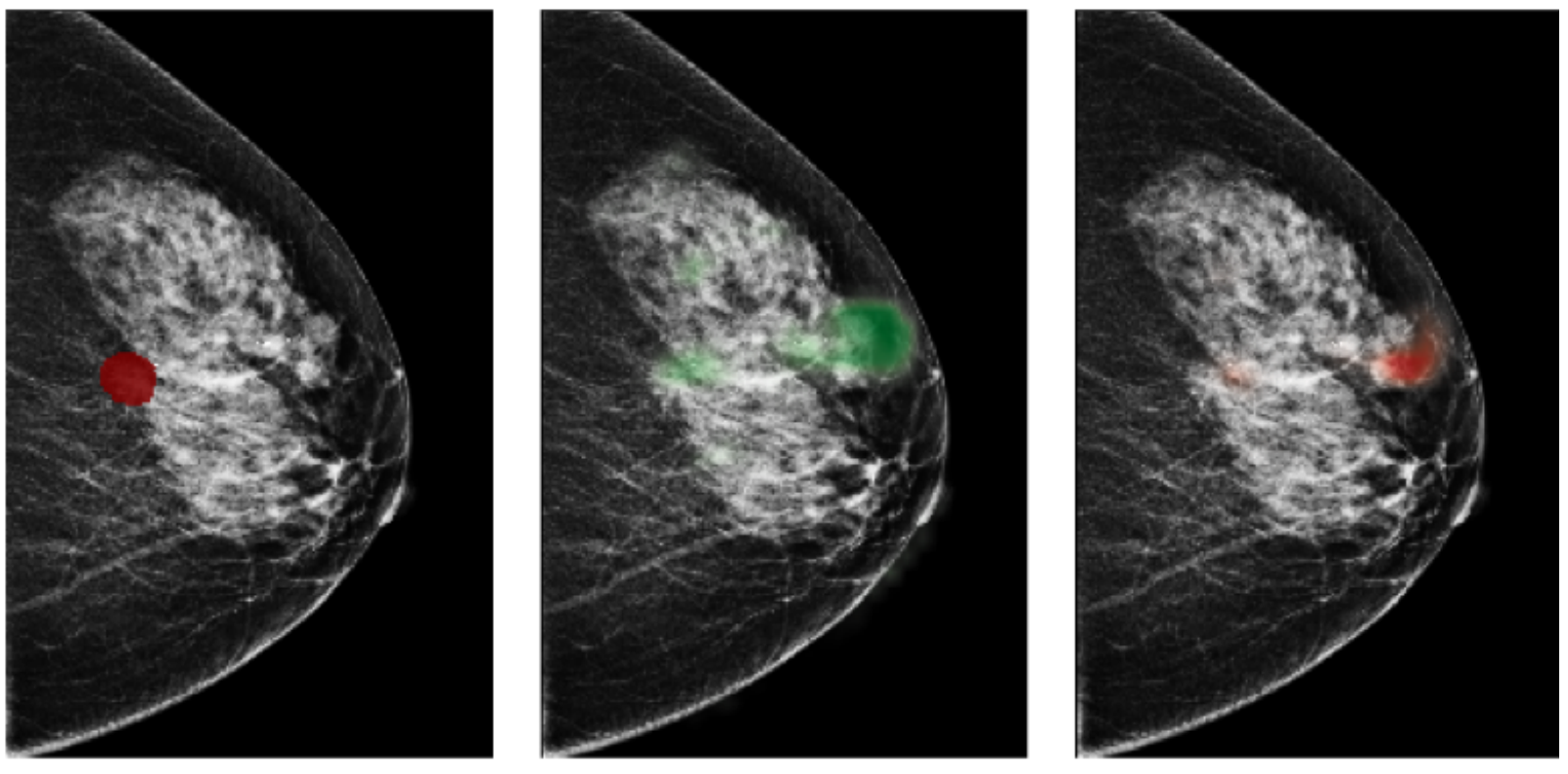}
    \end{tabular}
    \caption{An example of saliency maps. From left to right: a mammogram image of a right breast from craniocaudal view view (R-CC) with an annotated malignant lesion, a saliency map indicating suspicious regions for benign tissue, a saliency map indicating suspicious regions for malignant lesions.}
    \label{fig:saliency}
\end{figure}

In this study, we consider lesion-level classification, and design models to directly distinguish biopsy-confirmed lesions as being either benign or malignant. With this strategy, we enable the models to make accurate lesion-wise predictions. To show that deep learning approaches can benefit from utilizing global image context in classifying local findings on mammograms, we first train DNNs with cropped image patches to enable the learning of fine details from a specific region, then integrate the extracted local information with the global context. The global context is provided in the form of saliency maps (Figure~\ref{fig:saliency}) extracted by a model classifying the entire image. Here we use Globally-Aware Multiple Instance Classifier~\cite{shen2020interpretable} as the model to provide such saliency maps. In addition, we evaluate the models' performance on a challenging population which consists only of cases that are difficult to diagnose and the radiologist requested a biopsy for. This further differentiates our work from previous works~\cite{wu2019deep, shen2019globally, shen2020interpretable, kyono2018mammo, mckinney2020international} and makes our results not directly comparable to theirs. This is because these methods were developed and evaluated for the screening population, which contains a lot of negative cases not requiring biopsy, which can inflate their evaluation metrics~\cite{wu2019deep}.

Our results show that DNNs trained with image patches can effectively decrease the number of unnecessary biopsies, and that this ability can be further improved by utilizing global image context. Our best model is able to distinguish between benign and malignant findings on a test set of 600 lesions, achieving an area under the receiver operating characteristic curve (AUC) of 0.799$\pm$0.002.  If the model is utilized to assist in reading mammograms, about 2\% unnecessary biopsies could be avoided while catching all malignancies, in addition to cases that radiologists can easily exclude as benign or normal and not needing additional imaging. It reduces 15\% more unnecessary biopsies than the model using only local information, when under the same level of missed malignancies (2\%). It is worth noting that these performance measurements are computed on the population from which we exclude benign cases that radiologists can discount confidently by reading mammograms or other imaging exams. Overall, our results strongly suggest that the proposed strategy can be considered as a viable and valuable enhancer for deep learning methods in reducing unnecessary biopsies based on screening mammography.  

\section*{Materials and Methods}
This retrospective study was approved by our IRB and is compliant with the Health Insurance Portability and Accountability Act. Informed consent was waived.

\subsection*{Data}
We utilize a dataset consisting of 229,426 digital screening mammogramphy exams (1,001,093 images) from 141,473 unique patients screened between 2010 and 2017~\cite{wu2019nyu}. Each exam has four standard views and the resolution of images is approximately 2000${\times}$3000 pixels. We asked fellowship-trained breast imagers to annotate both benign tissues (e.g. cyst, fibroadenoma, fibrocystic change) and malignant lesions (e.g. IDC, ILC, DCIS), on the pixel-level. In the entire dataset, there are 8842 lesions from 8080 images with diagnosis confirmed by biopsy, which reveals the fact that a single breast can contain multiple lesions of differing types. The dataset is divided into disjoint training (80\%), validation (10\%) and test (10\%) sets. Detailed statistics of training, validation and test sets are in Table~\ref{tab:data_overall_stats}.

\begin{table}[ht]
\centering
\caption{Number of biopsy-confirmed lesions and number of mammogram images presenting no lesions, benign tissues, and malignant lesions in the training, validation and test set.}
\begin{tabular}{@{}lcccccc @{}}  
    \toprule
     & \multicolumn{3}{c}{\textbf{images}} &\phantom{a}&\multicolumn{2}{c}{\textbf{lesions/tissues}} \\
     \cmidrule{2-4} \cmidrule{6-7}
     & \textbf{negative} & \textbf{benign} & \textbf{malignant} &&\textbf{benign} & \textbf{malignant} \\
  \midrule 
   \textbf{training} & 808,730 & 5,188 &1,648 && 5,602 & 1,790\\
   \textbf{validation} & 123,130& 687 & 110 && 722 & 128\\
  \textbf{test} & 60,959& 432& 116& &464 & 136\\
  \midrule
  \textbf{overall} & 992,819 & 6,307&  1,874&&6,788&2,054 \\
    \bottomrule \\
\end{tabular}
\label{tab:data_overall_stats}
\end{table}

\subsection*{The proposed method}

Lesions in mammograms vary in size and shape, so if we crop these regions entirely and resize them to the same size to use them as inputs to standard deep neural networks, we will introduce information distortion and lose the fine details of the lesions. Therefore, we start by learning features of a number of image patches that are cropped from regions overlapping with the lesion, and then aggregate information from all patches to render a prediction for that lesion. 

\begin{figure}[htb!]
    \centering
        \begin{tabular}{cc}
        \multicolumn{2}{c}{\includegraphics[width=0.85\textwidth]{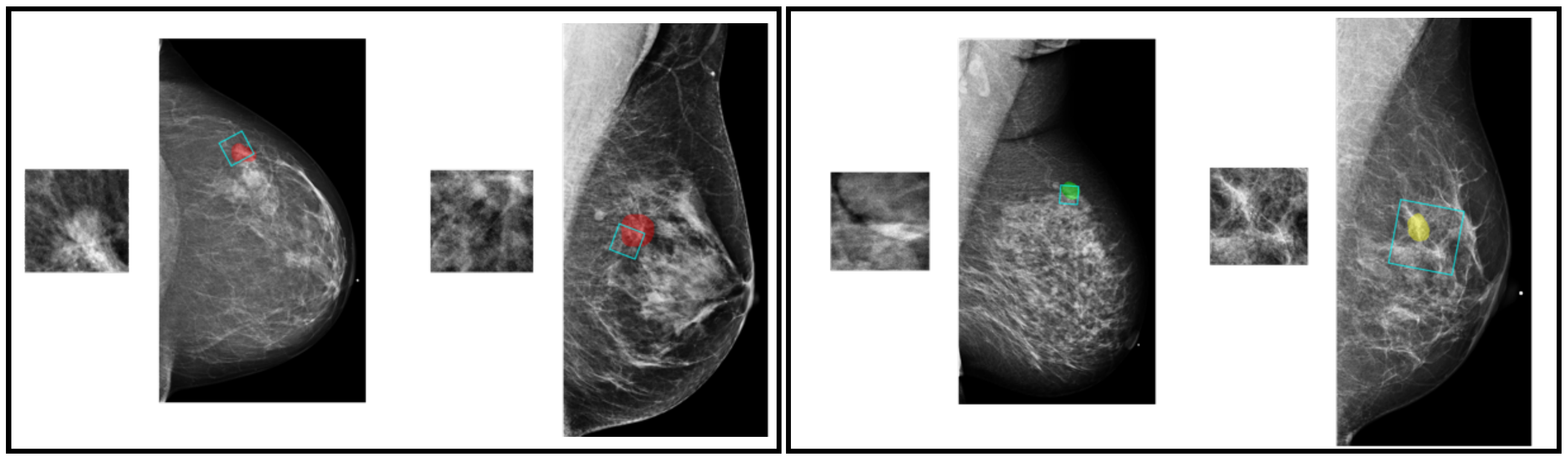}}\\
             \hspace{2mm}\textbf{a} & \textbf{b}  \\
             \multicolumn{2}{c}{\includegraphics[width=0.85\textwidth]{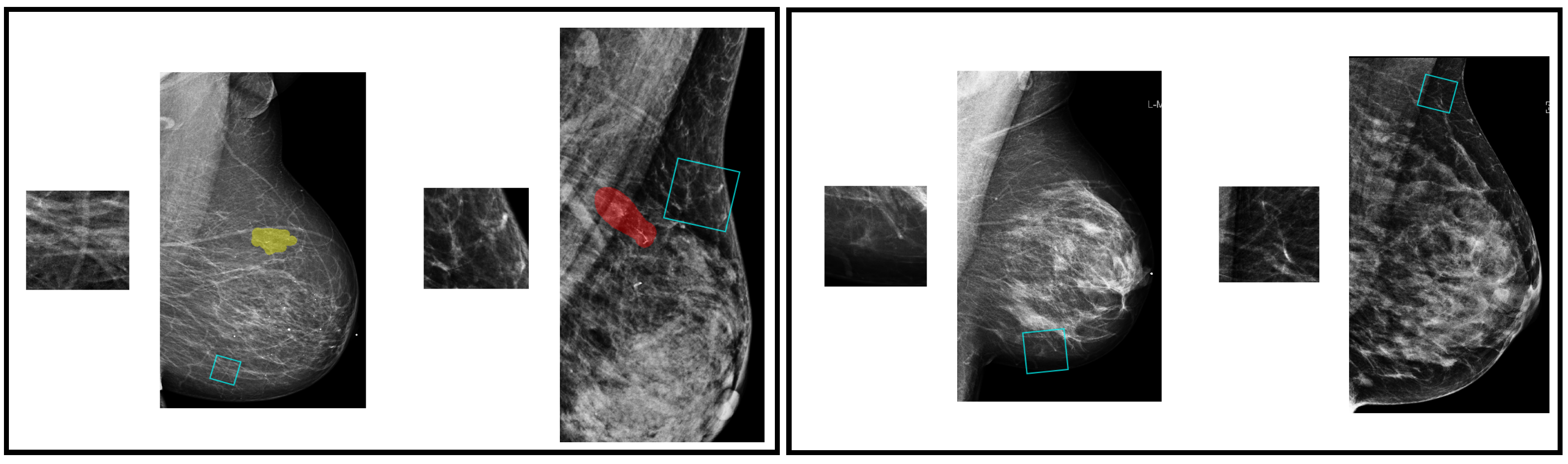}}\\
            \hspace{2mm}\textbf{c} & \textbf{d}
        \end{tabular}
    \caption{Examples of image patches along with the mammogram images from which they come. \textbf{a}. ``malignant'' patches, which overlap only with malignant findings (marked with red); \textbf{b}. ``benign'' patches, which  overlap only with benign findings (marked with yellow or green); \textbf{c}. ``outside'' patches, which are from regions outside the annotated lesions;  \textbf{d}. ``negative'' patches, which are from images without any biopsied findings. }
    \label{fig:patches}
\end{figure}

To extract information from image patches, we train a deep convolutional neural network (DCNN) to classify image patches of 256${\times}$256 pixels as one of the four classes: ``malignant'', ``benign'', ``outside'' and ``negative.'' Malignant and benign patches are cropped from windows that overlap with the segmentation of a malignant lesion or benign tissue. Besides cropping image patches that overlap with the annotations, we sample patches that have no overlap with any lesion (``outside''), as well as patches from breasts without records of biopsy (``negative''). The inclusion of these additional data is intended to regularize the model similarly to data augmentation. Examples of patches from each class are shown in Figure~\ref{fig:patches}.

\begin{figure}[htb!]
    \centering
    \begin{tabular}{cc}
        \includegraphics[width=0.26\textwidth]{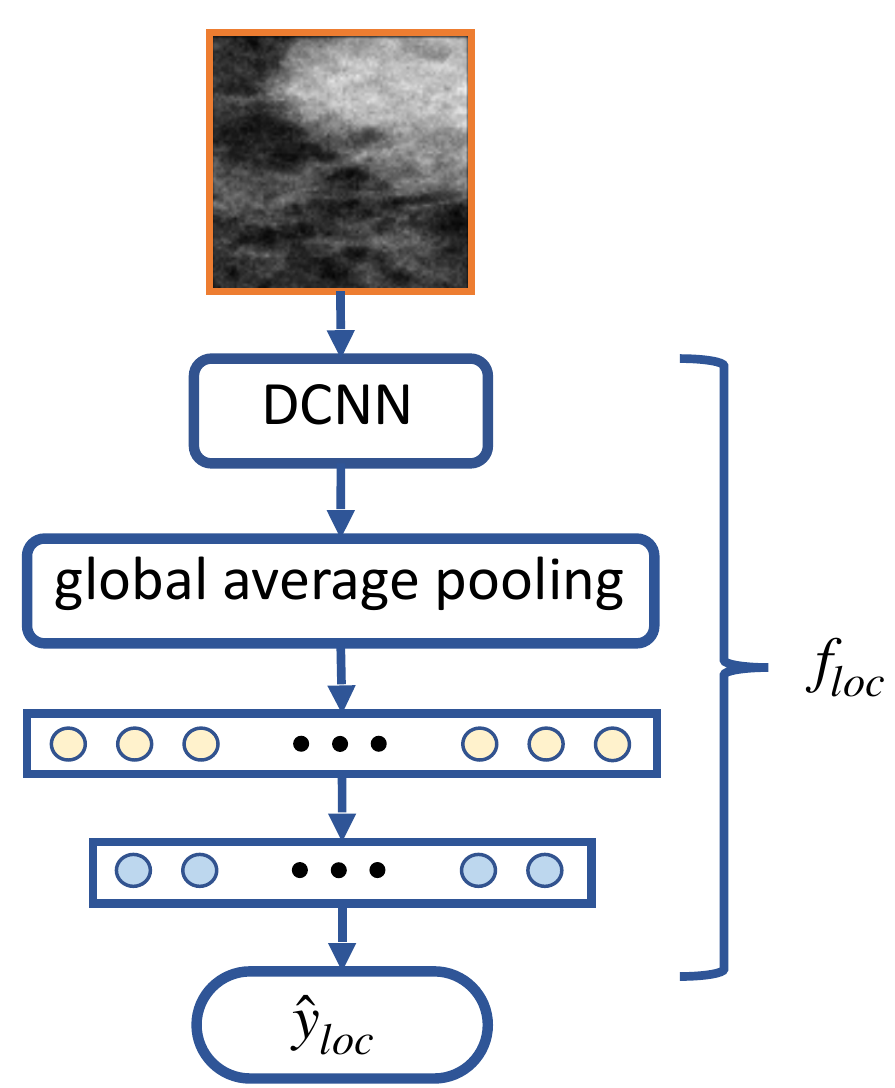} &\includegraphics[width=0.4\textwidth]{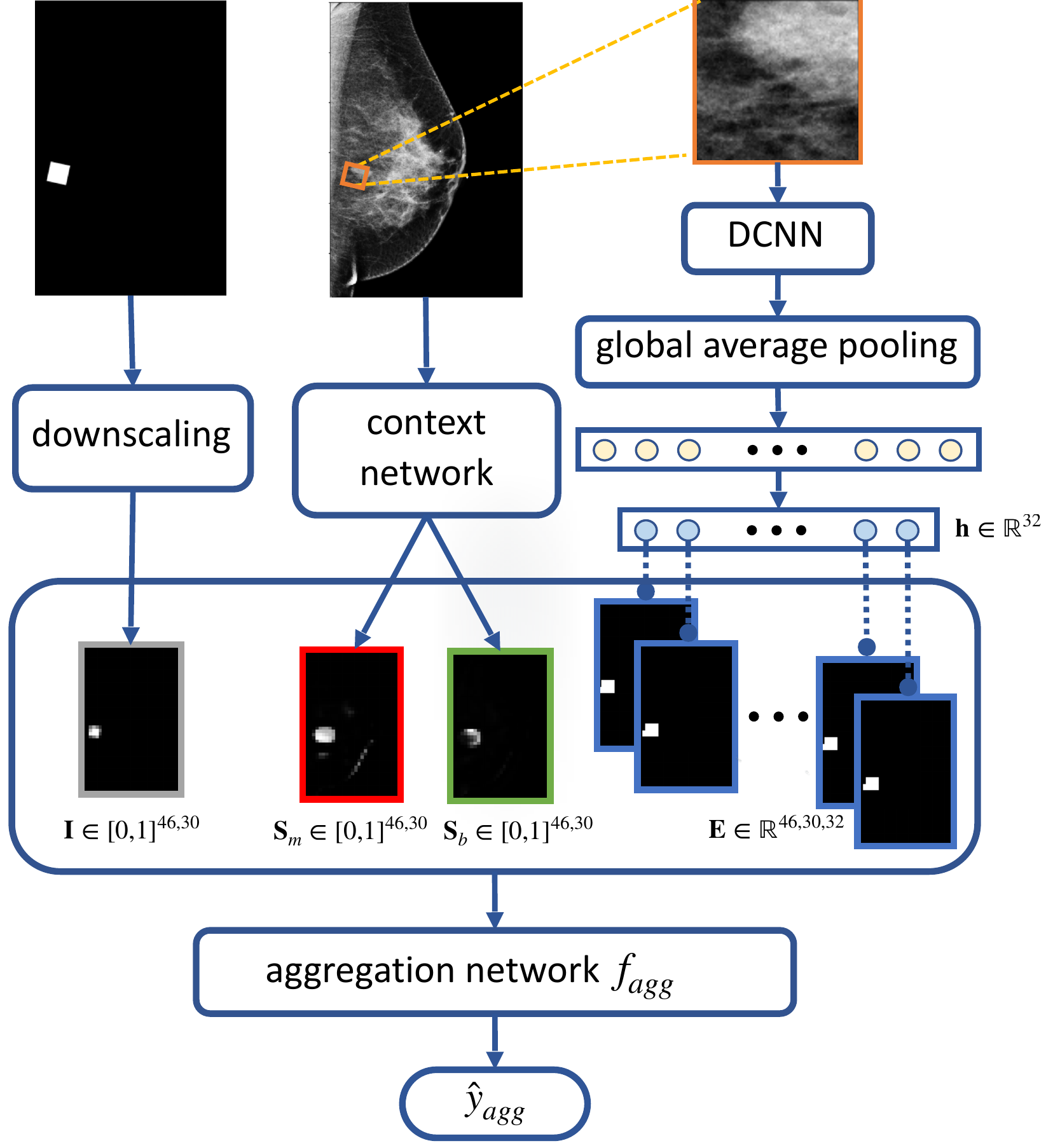}
    \end{tabular}
    \caption{Illustration of the proposed method. Left: a deep convolutional neural network which takes image patches of 256$\times$256 pixels as inputs, denoted as $f_{loc}$. Right: the aggregation network, $f_{agg}$, that takes the concatenation of three types of maps as inputs: 1) location indicator map, $\mathbf{I}$, in gray, generated by downscaling the binary mask indicating the cropping window's location, 2) saliency maps, $\mathbf{S}_m$ and $\mathbf{S}_b$, in red and green, generated by the context network, based on Globally-Aware Multiple Instance Classifier~\cite{shen2020interpretable}, 3) embedding map, $\mathbf{E}$, in blue, formed by the representation $\mathbf{h}$, produced by $f_{loc}$.}
    \label{fig:model}
\end{figure}

This DCNN is used to produce representations of local information. It is denoted as $f_{loc}$ and is shown in Figure~\ref{fig:model}a. We use DenseNet-161~\cite{huang2017densely} as its architecture. We add an additional fully-connected layer with 32 neurons between the global average pooling layer and the classification layer to obtain concise representations of the patch. The additional layer results in the feature vector $\mathbf{h} \in \mathbb{R}^{32}$ for the patch, which we use as the representation for the local information extracted by $f_{loc}$. 

To further incorporate global image context and curate the local information extracted by the DCNN, we train an ``aggregation network'' with inputs formed by aggregating maps containing information relative to the patch and the image it is cropped from, as illustrated in Figure~\ref{fig:model}b. This aggregation network is a shallow convolutional network, denoted as $f_{agg}$. It consists of two convolutional layers, each with 32 $3{\times}3$ convolutional filters, a global average pooling layer and, finally, a classification layer. We apply batch normalization and the ReLU activation function prior to each convolutional layer. This network is trained for the same patch classification task. The maps formed as inputs to the aggregation network are described in the following paragraphs.

The first type of maps are saliency maps which represent global context. We generate the saliency maps by training a network on full-resolution mammography images to predict the presence of benign and malignant lesions in the breast. We refer to this network as the ``context network.'' We use Globally-Aware Multiple Instance Classifier~\cite{shen2019globally, shen2020interpretable} as the context network, which is explicitly designed to provide interpretability by highlighting the most informative regions of the input images. To be more precise, the feature maps obtained after the last residual block of the context network are transformed by a $1{\times}1$ convolutional layer with sigmoid activation into two saliency maps, denoted as $\mathbf{S}_m$ $\in [0,1]^{46,30}$ and $\mathbf{S}_b$ $\in [0,1]^{46,30}$. Each pixel in the saliency map corresponds to a region in the full image, and its element denotes a score indicating the contribution of this region towards classifying the input image as containing malignant lesions or benign tissues. A pair of saliency maps for an image is shown in Figure~\ref{fig:saliency}.

Another type of maps are location indicator maps. Given a patch, this map indicates its cropping window's location on the mammogram, but is downscaled to be the same size as the saliency maps. The location indicator map is denoted as $\mathbf{I} \in [0,1]^{46,30}$. Same with the saliency maps, each pixel on the location indicator map corresponds to a region in the full image and the value of this pixel reflects how much the region is covered by the patch.

The last type of maps, called embedding maps, are formed by utilizing the representation $\mathbf{h}$ generated by $f_{loc}$. To construct this map, for each $\mathbf{h}_k \in \mathbf{h}$, we take a copy of the location indicator map of the patch $\mathbf{I}$ and replace its nonzero elements by $\mathbf{h}_k$. We denote the obtained map by $\mathbf{E}_k$. We concatenate $\mathbf{E}_k$'s to form the full embedding map $\mathbf{E} \in \mathbb{R}^{46, 30, 32}$. The embedding maps contain information learned by $f_{loc}$ specific to the fine details in the image patch.

These maps can be concatenated along the last dimension and served as inputs to the aggregation network, denoted as $\mathbf{X} \in \mathbb{R}^{46, 30, M}$ where $M$ is the number of maps. For example, when using both embedding maps and saliency maps as inputs, $M$ would be 34.   

\subsection*{Model training}

We first train $f_{loc}$ that takes image patches as inputs, followed by the aggregation network $f_{agg}$, which takes the concatenated maps, $\mathbf{X}$, as inputs. We use 20, 35, 5000, 4945 patches for malignant, benign, outside and negative patch classes in each training epoch. For data augmentation, we use random rotations (-30 to 30 degrees), and random sizes (128${\times}$128 to 384${\times}$384 pixels) when setting the cropping window to obtain the patch.

In order to address the extreme class imbalance, we use weighted cross-entropy as the training loss. The class weight for each patch class is set as inverse to the ratio of patches from this class among all patches used in each epoch. Therefore, losses on incorrect predictions of ``malignant'' and ``benign'' patches are appropriately up-weighted. For both $f_{loc}$ and $f_{agg}$, we adopt the same configuration while using $256{\times}256$ image patches as inputs for $f_{loc}$, and the concatenated maps, $\mathbf{X}$, as inputs for $f_{agg}$, in which the embedding maps is produced by the best performing $f_{loc}$ and saved to be used.

We minimize the training loss with the Adam optimizer~\cite{kingma2014adam}, setting the batch size to 25 for training $f_{loc}$, and 100 for $f_{agg}$. We initialize weights in $f_{loc}$ of all layers except the last two fully connected layers with weights from DenseNet-161~\cite{huang2017densely} pretrained on the ImageNet ILSVRC-2012 dataset~\cite{russakovsky2015imagenet}, then fine-tune the entire network. We randomly initialized the weights of $f_{agg}$. We optimize the hyperparameters using random search~\cite{bergstra2012random}. Specifically, we search for the learning rate on the logarithmic scale in $[10^{-6}, 10^{-4}]$ for $f_{loc}$, and in $[10^{-5}, 10^{-3}]$ for $f_{agg}$. Early stopping is performed if we observed that the AUC on the validation set has not increased for ten epochs. We implement the models in PyTorch~\cite{paszke2019pytorch}, and use NVIDIA Tesla V100 GPUs for model training and inference. 

\subsection*{Model evaluation}
\begin{figure}[htb!]
    \centering
        \begin{tabular}{cc}
        \multicolumn{2}{c}{\includegraphics[width=0.85\textwidth]{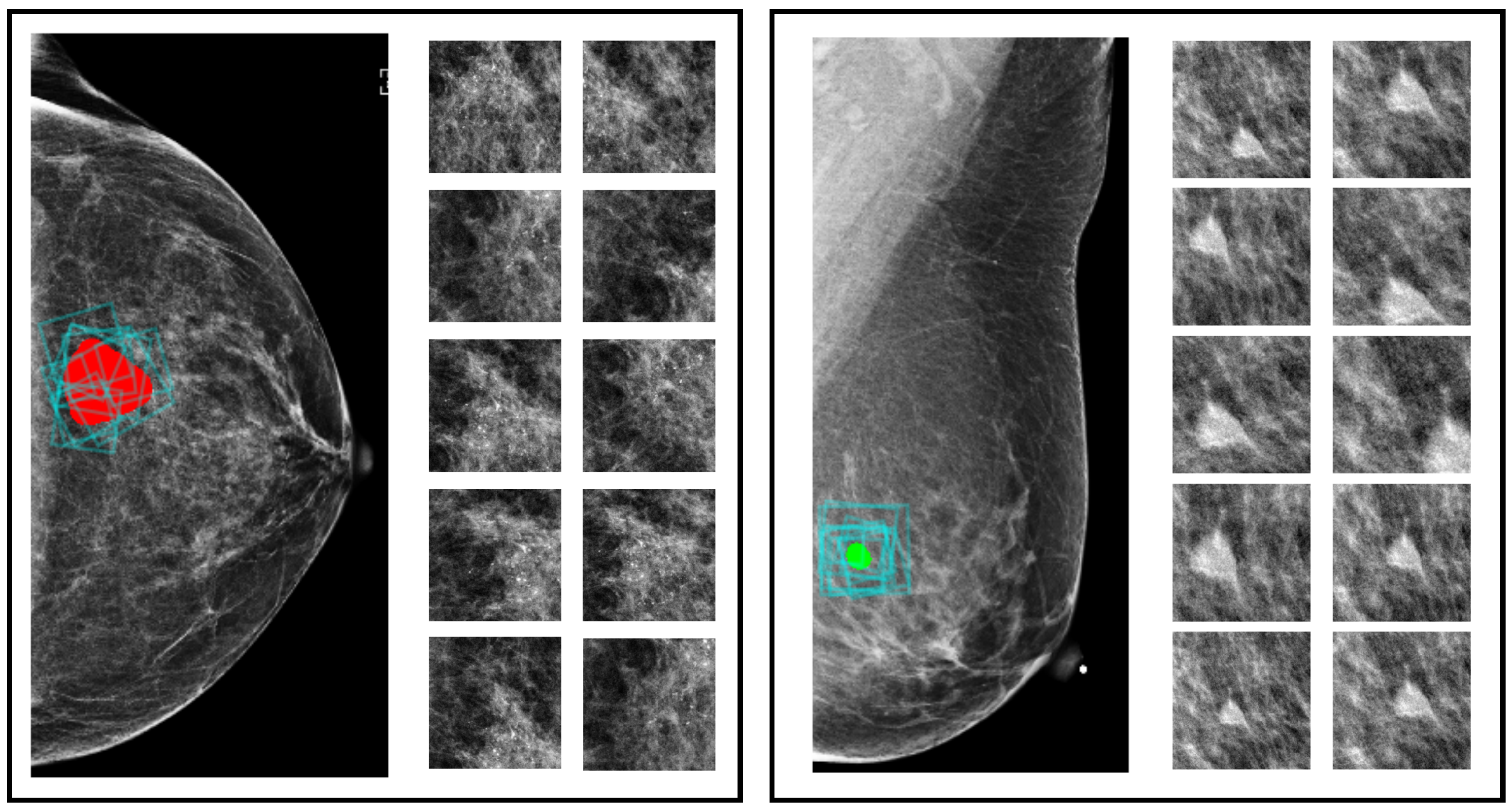} }\\
        \end{tabular}
    \caption{Multiple patches are sampled to obtain a prediction for one lesion. In each black box, we present a biopsied finding from the test set and ten out of 100 patches used by the model to make a prediction for the finding. Lesions are marked in red (malignant) or green (benign) and the cropping windows of the patches are marked by blue boxes on the image. Cropped patches are shown on the right.}
    \label{fig:evaluation}
\end{figure}

During training, we consider patches from mammograms with and without lesions, and perform multi-class classification over four classes: malignant, benign, outside and negative. In the validation and test phases, we only consider patches from images with lesions, and transform the patch-level predictions into a malignancy prediction for each lesion in the images. To get a prediction for a lesion, as shown in Figure~\ref{fig:evaluation}, we crop 100 patches that overlap with the segmentation of the lesion. The size of the cropping window varies from 128${\times}$128 to 384${\times}$384 pixels, which is the same range we used for data augmentation. After cropping, each patch is resized to 256${\times}$256 pixels, and we use it as input to $f_{loc}$ to produce a feature vector. Then, we apply $f_{agg}$ on the concatenated maps, including the embedding maps transformed by the feature vector and we get its prediction, each as four scores for the four patch classes. For each patch, we normalize the scores for malignant and benign patch classes so that they sum to one. Finally, we average the 100 normalized scores of the 100 sampled patches to obtain a prediction for the lesion. Based on these estimated probabilities, we compute the AUC that the model achieves in classifying the lesions as malignant or benign. We use the AUC computed on the 850 lesions from the validation set for model selection, and report the AUC computed for the 600 lesions from the test set.

\section*{Results}

\subsection*{Model performance}
We report the model's performance on the test set consisting of 600 lesions which are present on 534 images from 260 patients. There are 44 images containing more than one lesion, and 14 images have both benign and malignant lesions. We emphasize that if we were to use deep learning methods that provide only breast-level risk estimation, it would be impossible to tell which lesion is the one with higher risk of malignancy.

Both model components, $f_{loc}$, which takes patches as inputs, and the aggregation network $f_{agg}$, are selected according to their performance on the validation set. The best performing aggregation network using embedding maps and saliency maps achieved an AUC of 0.799$\pm$0.002 on the test set. In Table~\ref{tab:fnr}, we include more results on the portion of unnecessary biopsies that can be avoided while missing a given portion of malignancies when using the model’s prediction as a second reader to assist radiologists. It can help to further reduce 1.7\% unnecessary biopsies in addition to cases that radiologists can easily exclude as benign or normal and not needing additional imaging, while catching all malignancies. According to the estimated yearly cost related with unnecessary biopsies~\cite{vlahiotis2018analysis, ong2015national}, it can be translated to saving more than a million dollars each year in the US for breast-care. If reducing biopsies is prioritized further, up to 13.5\% could be avoided while only missing 1\% of malignancies and up to 23.1\% could be avoided while only missing 2\% of malignancies.

\begin{table}[htb!]
    \centering
    \caption{True negative rate (TNR) achieved by our model when its false negative rate (FNR) are 0.01, 0.02, 0.03 and 0.05 as we vary prediction threshold for assigning observations to a positive class indicating malignancy. The 95\% confidence interval of the estimated TNR and the clinical meaning are also provided.}
    \begin{tabular}{@{}cccc@{}}  
    \toprule 
      clinical meaning & FNR & TNR &  95\% CI of TNR \\
  \midrule 
     \begin{tabular}{c}
      1.7\% unnecessary biopsies we can help to avoid \\
while missing no malignancies 
 \end{tabular} & 0.00 & 0.017 & $[-0.005, 0.040]$\\
   \begin{tabular}{c}
      13.5\% unnecessary biopsies we can help to avoid \\
while missing 1\% malignancies 
 \end{tabular} & 0.01 & 0.135 & $[0.039, 0.232]$\\
 
 \begin{tabular}{c}
      23.1\% unnecessary biopsies we can help to avoid \\
while missing 2\% malignancies 
 \end{tabular} & 0.02 & 0.231 & $[0.190, 0.273]$\\
 
  \begin{tabular}{c}
      27.6\% unnecessary biopsies we can help to avoid \\
while missing 3\% malignancies 
 \end{tabular} & 0.03 & 0.276 & $[0.240, 0.312]$\\
 
   \begin{tabular}{c}
      43.5\% unnecessary biopsies we can help to avoid \\
while missing 5\% malignancies 
 \end{tabular} & 0.05 & 0.435 & $[0.407, 0.562]$\\
 
    \bottomrule 
\end{tabular}

    \label{tab:fnr}
\end{table}

\subsection*{Ablation experiments} 
We conduct the following experiments to justify the choice of model architecture, to verify impact of transfer learning, and to elaborate the importance of utilizing both local fine details and global image context in identifying malignant lesions.

\begin{figure}[htb!]
    \centering
        \includegraphics[width=0.75\textwidth]{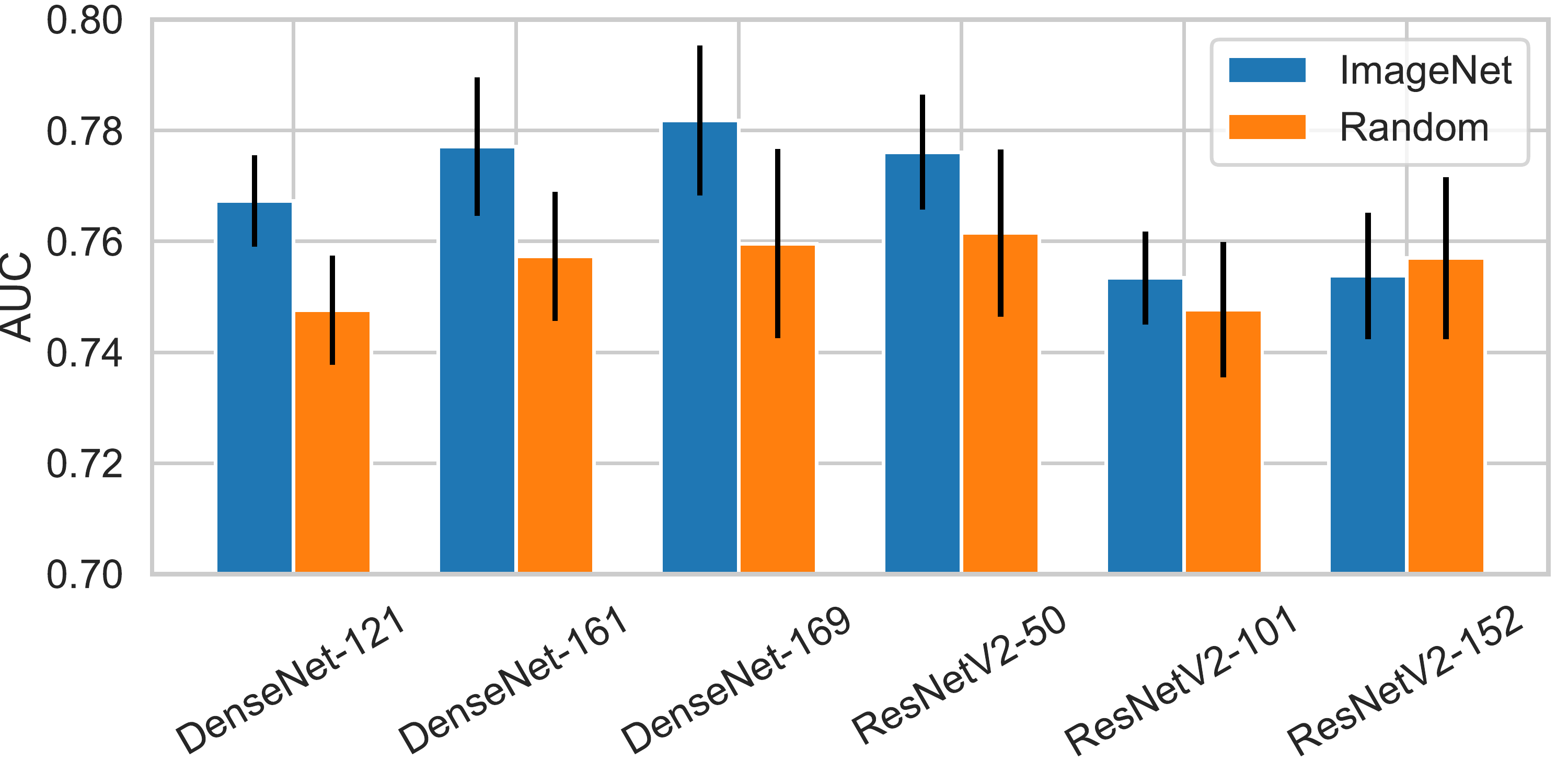} 
    \caption{Test performance in classifying biopsied findings, achieved by $f_{loc}$, when using different architectures and weights initialization strategies. }
    \label{fig:perform}
\end{figure}

\paragraph{Architecture search.}
To choose the architecture for $f_{loc}$, we considered a number of ResNet~\cite{he2016deep} and DenseNet~\cite{huang2017densely} variants. These architectures use skip connections, which improve information flow between layers, and allow for effective training of very deep networks. They both achieved strong results across a wide range of image classification tasks~\cite{guan2020multi, wang2017chestx, hannun2019cardiologist}. The specific ResNet and DenseNet variants we experimented with are: ResNetV2-50, ResNetV2-101, ResNetV2-152, DenseNet-121, DenseNet-161, and DenseNet-169. We compared the performance of $f_{loc}$ when being parameterized as the above architectures. The results are shown in Figure~\ref{fig:perform}. In this experiment, we did not consider global context and used the predictions made by $f_{loc}$ for each lesion. 

\paragraph{Transfer learning.}
Transfer learning by pretraining the network on a different task is widely adopted to improve neural networks' performance. We experiment with initializing $f_{loc}$ with weights from networks pretrained on the ImageNet ILSVRC-2012 dataset~\cite{russakovsky2015imagenet}, and compare it to initializing the weights randomly using He initialization~\cite{he2015delving}. Since images from ImageNet are RGB while mammograms are grayscale, we duplicated each patch across the three channels. The AUCs achieved by $f_{loc}$ with or without using transfer learning are presented in Figure~\ref{fig:perform}. Without transfer learning, ResNetV2-50 achieved the highest AUC of 0.762$\pm$0.015, while DenseNet-121 achieved the lowest AUC of 0.748$\pm$0.098. When we applied transfer learning, the performance is improved for most of the architectures except ResNetV2-152, and DenseNet-169 becomes the best performer with 0.782$\pm$0.014 AUC. We conclude from these results that transfer learning from the ImageNet dataset~\cite{russakovsky2015imagenet} clearly improves our results, even though the natural image domain and the mammography image domain are so different.

\paragraph{Importance of global context.}

To assess the importance of global context in classifying lesions localized to small regions of the image, we performed further ablation experiments. We trained networks using different combinations of saliency maps, location indicator map and embedding maps as inputs. Selected maps are concatenated along the last dimension and used by the aggregation network. Table~\ref{tab:combin} presents the results when using all possible combinations. Since ImageNet-pretrained DenseNet-161 as $f_{loc}$ achieved the highest AUC on the validation set, we used it in this set of experiments. 

As expected, for the task of classifying biopsy-confirmed lesions, most of the predictive power comes from local features: the aggregation network trained with only embedding maps achieved an AUC of 0.778$\pm$0.002. In comparison, the network trained only with saliency maps achieved an AUC of 0.695$\pm$0.003, indicating that global context alone was not highly predictive. When we introduced location indicator maps together with saliency maps into the network, the AUC increased to 0.721$\pm$0.011. We observed that using location indicator maps and embedding maps together does not improve performance. This is unsurprising since embedding maps contain the same location information conveyed by the location indicator maps. Finally, networks using the combination of embedding maps and saliency maps achieved an AUC of 0.799$\pm$0.002, which is the highest among all combinations. The fact that combining local features with global context outperformed each of them in isolation confirms the importance of the global image context in classifying lesions localized to small regions of the mammogram. 

\begin{figure}[htb!]
    \centering
        \includegraphics[width=0.75\textwidth]{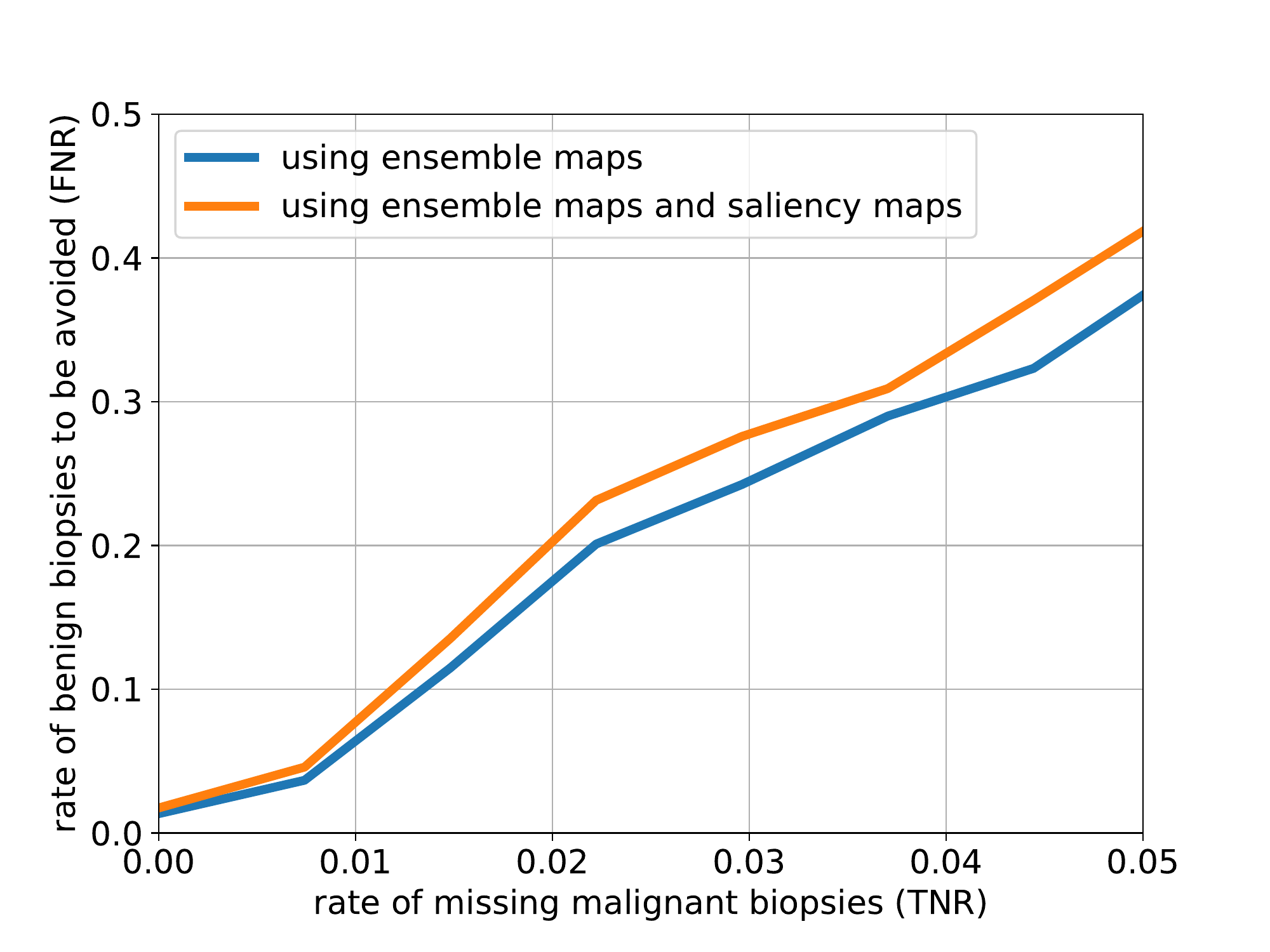} 
    \caption{True negative rate (TNR) and false negative rate (FNR) achieved by the aggregation network using ensemble maps or using both ensemble maps and saliency maps as we vary the prediction threshold for assigning observations to a given class.}
    \label{fig:fnr}
\end{figure}

To further evaluate the impact of the global context in our task, we investigate the relation between the true negative rate (indicating biopsies that can help to avoid) and the false negative rate (indicating missed malignancies). Specifically, we compare the aggregation network using only the ensemble maps and the aggregation network using both the ensemble maps and the saliency maps. Figure~\ref{fig:fnr} visualizes this relationship. We considered scenarios when there are less than 5\% malignancies missed. For all considered false negative rates utilizing the saliency maps resulted in lower true negative rate.

\begin{table}[htb!]
    \centering
    \caption{Test performance of the aggregation network when using different information combinations as inputs. Models utilizing both local and global information achieved better performance than the counterparts using single type of maps.}
    \begin{tabular}{@{}lc@{}}  
    \toprule 
      & AUC \\
  \midrule 
   location indicator maps & 0.474 $\pm$ 0.031\\
   embedding maps & 0.778 $\pm$ 0.002\\
    saliency maps & 0.695 $\pm$ 0.003\\
    location indicator maps + embedding maps & 0.777 $\pm$ 0.002\\
      location indicator maps + saliency maps & 0.721 $\pm$ 0.011\\
       embedding maps + saliency maps& \textbf{0.799 $\pm$ 0.002}\\
        location indicator maps + saliency maps + embedding maps & 0.797 $\pm$ 0.001\\
    \bottomrule 
\end{tabular}

    \label{tab:combin}
\end{table}

\section*{Discussion}

Regular screening mammography is widely acknowledged to be the best way to detect breast cancer early. However, mammogram-based diagnosis performed by radiologists suffers from a high false positive rate, resulting in both unnecessary imaging and tissue biopsies. Developing deep learning technologies to assist breast cancer screening is promising, but previous works in the literature rarely focused on reducing unnecessary biopsies. Besides achieving radiologist-level performance at detecting breast cancer in mammograms, deep learning models are expected to play a more important role in distinguishing whether a given lesion is malignant or benign. This distinction is highly beneficial for the case of suspicious-appearing but ultimately benign findings that result in unnecessary biopsies by the radiologist.

In this study, we presented a method to combine local features in small image patches with global context in high resolution mammogram images. We showed that it is necessary to consider both fine details in a small region and the global image context to improve deep learning models' performance when classifying localized lesions on the high resolution images, while previous works usually consider only image patches or downscaled mammogram images~\cite{xi2018abnormality, spanhol2016breast, aboutalib2018deep}. Our resulting deep learning model achieved an AUC of $0.799\pm0.002$ in classifying biopsy-confirmed lesions as being malignant or benign. It can help to further reduce over 23\% of unnecessary biopsies while missing only 2\% of cancer as the second reader on regions that radiologists have low confidence on. Compared with works performing breast-level classification~\cite{aboutalib2018deep, wu2019deep, elter2007prediction, xi2018abnormality, shen2019globally, shen2020interpretable, mckinney2020international}, our model can provide prediction for each individual suspicious lesion, and therefore present precise guidance for followup procedure including biopsy and surgery.

We acknowledge some limitations of this work. For instance, we did not capture the levels of difficulty related to different types of cancer, which is clinically valuable. We leave this for future work. In addition, the context network we considered in the study did not perform cross-view reasoning, and we expect that networks utilizing all four standard views in a mammogram exam can introduce more complete information and result in more reliable cancer detection.

\section*{Conclusion}
Besides performing breast-level classification, deep learning methods can help further reduce unnecessary biopsies by classifying suspicious small regions as being benign or malignant. Furthermore, incorporating global image context can improve the network's ability to distinguish between localized benign and malignant lesions on high resolution images. Future research on techniques for combining local information with global context may be promising for breast cancer screening.

\bibliographystyle{custom}
\bibliography{sample}

\end{document}